\documentclass[12pt]{iopart}
\usepackage{bm}
\usepackage{hyperref}

\renewcommand{\epsilon}{\varepsilon}
\begin{document}
\title{Cosmic No Hair for Collapsing Universes}

\author{James E. Lidsey} 
\address{Astronomy Unit, School of Mathematical Sciences\\
  Queen Mary, University of London\\
  Mile End Road, London E1 4NS\\
  United Kingdom}
\eads{\mailto{J.E.Lidsey@qmul.ac.uk}}

\begin{abstract}
It is shown that all contracting, spatially homogeneous, orthogonal 
Bianchi cosmologies that are
sourced by an ultra-stiff fluid with an arbitrary and, in general, 
varying equation of state asymptote to the spatially flat and isotropic 
universe in the neighbourhood of the big crunch singularity. 
This result is employed to investigate the asymptotic dynamics of 
a collapsing Bianchi type IX universe sourced by a scalar field 
rolling down a steep, negative exponential potential. 
A toroidally compactified version of ${\rm M}^*$-theory that leads to 
such a potential is discussed and it is shown that the isotropic 
attractor solution for a collapsing Bianchi type IX universe
is supersymmetric when interpreted in an eleven-dimensional context. 
\end{abstract}

\maketitle

 


\section{Introduction}
\setcounter{equation}{0}

The possibility that our universe underwent a `pre-big bang' 
contraction before bouncing into its present expansionary 
phase continues to attract attention (see, e.g., 
\cite{ven,lidsey,cyclic}). In scenarios
of this type, a central question to address is the  
behaviour of the universe during the final stages of 
the collapse in the vicinity of the big crunch singularity. 
Recently, Erickson {\em et al.} \cite{ewst} have considered the 
class of collapsing, spatially homogeneous cosmologies
that are sourced by an ultra-stiff perfect 
fluid with an equation of 
state $p= (\gamma  -1 ) \rho $, where $\gamma >2$. 
In the case of a constant equation of state parameter, 
they have proved that the universe asymptotes to the 
spatially flat and isotropic 
Friedmann-Robertson-Walker (FRW) cosmology on the approach to the big crunch.  
This is to be expected on qualitative grounds, since 
the energy density of the matter source scales as 
$\rho \propto a^{-3\gamma}$, whereas the curvature 
and anisotropies vary as $a^{-2}$ and $a^{-6}$, respectively. The latter
therefore become subdominant as the spatial volume $a^3 \rightarrow 0$. 
More generally, for a variable equation of state, they have argued 
that if the spatial curvature and anisotropy are initially small, 
they remain so during the collapse. 

In this paper we provide a rigorous proof of a `cosmic no hair' result
for all contracting, orthogonal, spatially homogeneous Bianchi cosmologies,
where the matter source has an arbitrary 
equation of state $\gamma =\gamma (\rho )$ that is 
subject only to the condition 
that it is differentiable and bounded such that 
$(\gamma -2 )$ is positive-definite for all time (including at
the big crunch). One example of a matter source with an 
equation of state $\gamma > 2$ is a minimally coupled 
scalar field with a negative-definite self-interaction potential.  
We employ the cosmic no hair theorem to gain insight 
into the behaviour of a collapsing Bianchi type IX 
cosmology sourced by a scalar field interacting through  
a steep, negative exponential potential. 
We find that such a universe can indeed isotropize at the big crunch. 

Negative exponential 
potentials are known to arise in compactifications 
of higher-dimensional theories such as ${\rm M}^*$-theory \cite{hull}. 
This is an eleven-dimensional theory in a spacetime 
with signature $(9+2)$ and is directly related to 
M-theory \cite{witten} by string dualities. For a particular 
toroidal compactification 
of this theory to four dimensions, we interpret the isotropic 
attractor solution 
for a contracting Bianchi type IX cosmology in an 
eleven-dimensional context and find that it corresponds to a
supersymmetric background.   

\section{Cosmic No hair Theorem}

\subsection{Bianchi Cosmology}

Bianchi models are spatially homogeneous cosmologies admitting a 
three--parameter local group $G_3$ of isometries that acts simply
transitively on spacelike hypersurfaces $\Sigma_t$. 
Coordinates can be chosen such that the four--dimensional 
line element has the form $ds^2 =-dt^2 +h_{ab}(t) \omega^a \omega^b$ 
$(a,b=1,2,3)$, where the one--forms $\omega^a$ satisfy 
the Maurer--Cartan equation $d\omega^a= \frac{1}{2} {C^a}_{bc} 
\omega^b \wedge \omega^c$ and ${C^a}_{bc}$ are the structure 
constants of the Lie algebra of $G_3$. Since ${C^a}_{(bc)} =0$, 
${C^a}_{bc}$ has at most nine independent 
components and these are classified in terms of 
a symmetric $3\times 3$ matrix, $n^{ab}$, and the components 
of a $3 \times 1$ vector $A_b \equiv {C^a}_{ab}$. This implies that
the structure constants can be expressed in the form  
\begin{equation}
\label{defm}
{C^c}_{ab} \equiv n^{cd} \epsilon_{dab} + \delta^c_{[a} A_{b]} ,
\end{equation}
where $\epsilon^{abc}$ is the totally antisymmetric tensor 
with $\epsilon^{123}=1$. Substitution of  Eq. (\ref{defm}) into 
the Jacobi identity ${C^e}_{d[a} {C^d}_{bc]} =0$ then implies 
that $A_b$ is transverse to $n^{ab}$: 
\begin{equation}
\label{transverse}
n^{ab}A_b =0  .
\end{equation}
If $A_b \ne 0$, it represents an eigenvector 
of $n^{ab}$ with zero eigenvalue and  
it may be assumed without loss of generality 
that $n^{ab}={\rm diag} [ n_1, n_2, n_3 ]$ and that $A_b= (A, 0, 0 )$. 
Moreover, a suitable rescaling can always be found such that   
the eigenvalues of $n^{ab}$ take values $\{ 0, \pm 1\}$. As a result,  
Eq. (\ref{transverse}) simplifies to 
\begin{equation}
\label{simpletransverse}
n_1 A =0   .
\end{equation}
Bianchi class A models satisfy $A=0$ and the 
class B are defined by the property $A\ne 0$ $(n_1 =0)$ \cite{ellismac}.

Throughout this paper, we consider orthogonal Bianchi models
where the fluid velocity vector is orthogonal to the group orbits.  
In the case of a scalar field matter source, this implies that 
the field is constant on the surfaces of homogeneity. 
We employ the orthonormal frame approach developed by Wainwright 
and collaborators \cite{wainA,wainB}. (For a review see \cite{wainellis}).  
In this approach, the Hubble scalar is defined by $H\equiv \dot{\ell}/\ell$, 
where $\ell$ is a length scale and a dot denotes differentiation with 
respect to $t$. The Raychaudhuri and Friedmann equations are then given by  
\begin{eqnarray}
\label{Ray}
\dot{H}=-H^2-\frac{2}{3} \sigma^2- \frac{1}{6} (3\gamma -2) \rho \\
\label{Fried}
3 H^2 = \sigma^2 -\frac{1}{2} {^{(3)}}R +\rho  ,
\end{eqnarray}
respectively, where $\sigma^2 \equiv \frac{1}{2} \sigma_{ab}\sigma^{ab}$
is defined in terms of the shear tensor $\sigma_{ab}$, 
${^{(3)}}R$ is the scalar curvature 
of the $t={\rm constant}$ hypersurfaces, and $\rho$ and $p$ 
represent the energy density and pressure of the fluid.  
The Friedmann constraint equation (\ref{Fried}) may be expressed 
in the form 
\begin{equation}
\label{genfried}
\Omega + \Sigma^2 + K =1
\end{equation}
by defining expansion-normalized 
density, shear and curvature parameters:  
\begin{equation}
\label{defpara}
\Omega \equiv \frac{\rho}{3H^2}, \quad \Sigma^2 \equiv \frac{\sigma^2}{3H^2}
, \quad K= - \frac{{^{(3)}}R}{6H^2}  .
\end{equation}

A deceleration parameter,  $q \equiv 
- \ddot{\ell}\ell /\dot{\ell}^2$, may also be defined such that 
\begin{equation}
\label{dotH}
\dot{H}=-(1+q)H^2
\end{equation}
and Eqs. (\ref{Ray}) and (\ref{dotH}) together imply that 
\begin{equation}
\label{defq}
q=2\Sigma^2+\frac{1}{2} (3\gamma -2) \Omega  .
\end{equation}

Finally, the covariant 
conservation of energy-momentum is expressed in the form of the 
fluid equation:  
\begin{equation}
\label{fluid}
\dot{\rho} = - 3H \gamma \rho  ,
\end{equation}
where $p \equiv [\gamma (\rho ) -1 ] \rho$ defines the 
equation of state parameter, $\gamma (\rho)$. 
We will consider an arbitrary equation of state subject only to the 
conditions that $p=p(\rho )$ is 
at least ${\cal{C}}^1$ and that $(\gamma -2 )$ 
is positive-definite for all values of the 
spatial volume. We also assume the standard energy conditions 
hold and, in particular, that $\rho \ge 0$. 

The scalar curvature satisfies 
${^{(3)}}R \le 0$ for all Bianchi types except the
type IX. It then follows from the Friedmann equation (\ref{Fried}) 
that the spatial volume of an initially contracting  
Bianchi type I-VIII universe will decrease monotonically 
with cosmic time, $t$. In view of this, 
it proves convenient to define a dimensionless time variable $\tau$
\cite{wainA,wainB}: 
\begin{equation}
\label{deftau}
\frac{dt}{d\tau} \equiv  \frac{1}{H}  .
\end{equation}
Hence, $\tau$ is a monotonically decreasing function of $t$ when 
$H<0$ and, since $0<\ell <\infty$, the big crunch singularity 
will occur at $\tau \rightarrow -\infty$. 

We now show that $\Omega \rightarrow 1$ in the vicinity of the big 
crunch for all Bianchi type I-VIII universes. Firstly, 
Eq. (\ref{genfried}) implies that $\Omega \le 1$ since $K \ge 0$.  
Furthermore, Eqs. (\ref{genfried}), (\ref{defpara}), (\ref{dotH})  
and (\ref{fluid}) yield an evolution equation for the density parameter 
which takes the form: 
\begin{equation}
\label{Omegaevolve}
\Omega' = \left[ - (3\gamma -2) K +3(2- \gamma ) \Sigma^2 \right]
\Omega  ,
\end{equation}
where a prime denotes $d/d\tau$. It follows from Eq.   
(\ref{Omegaevolve}) that $\Omega' \le 0$ for any initially 
contracting Bianchi type I-VIII universe, with equality 
iff $K=\Sigma^2=0$ for any non--vacuum orbit $(\Omega >0)$. Hence, 
$\Omega$ is a monotonic decreasing function of $\tau$
and, since $\Omega$ is bounded, we may conclude that  
$\lim_{\tau \rightarrow -\infty} \Omega' = 0$. 
Eq. (\ref{Omegaevolve}) then implies that   
\begin{equation}
\label{limitKSO}
\lim_{\tau \rightarrow -\infty} K = \lim_{\tau \rightarrow -\infty} 
\Sigma =0 , \qquad \lim_{\tau \rightarrow -\infty} \Omega =1  ,
\end{equation}
where the latter limit follows directly from Eq. (\ref{genfried}). 

To proceed further, we will 
require the specific form of the Einstein field equations for each 
Bianchi model. For the Bianchi type I-VIII universes, these equations  
can be expressed in the form of an autonomous set of 
ordinary differential equations (ODEs) of the form 
${\bf x}'={\bf f} ({\bf x})$, subject to a constraint 
equation $g({\bf x})=0$, where the state vector 
${\bf x} \in \Re^6$. The physical interpretation 
of the state variables ${\bf x}$ is different for the class A and B models, 
however, and we therefore consider each class in turn in the following 
Subsections. 

\subsection{Bianchi Class A (I - VIII)}

The physical state of a Bianchi class A cosmology  
is determined by the vector
${\bf x} = (H, \Sigma_+, \Sigma_- , N_1,N_2,N_3)$, where 
$N_a\equiv n_a/H$, $\Sigma_{\pm} = \sigma_{\pm}/H$ and 
$\sigma_{\pm}$ are linear combinations 
of the two independent components of the (traceless) shear tensor
\cite{wainwright,wainA}. Thus, 
$\Sigma_{\pm}$ determine the anisotropy associated with the Hubble 
flow and are related to the shear parameter by 
\begin{equation}
\label{defSigma}
\Sigma^2= \Sigma^2_+ + 
\Sigma_-^2  ,
\end{equation}
whereas $N_a$ parametrize the spatial curvature of the group orbits
and are related to the curvature parameter such that 
\begin{equation}
\label{defK}
K=\frac{1}{12} \left[ N_1^2+N_2^2+N_3^2 -2(N_1N_2 +N_2N_3+N_3N_1)  
\right] .
\end{equation}

The Einstein field equations for the Bianchi class A 
are then given by an autonomous 
set of first-order ODEs \cite{wainA}:
\begin{eqnarray}
\label{efeA1}
\Sigma_{\pm}'= -(2-q)\Sigma_{\pm}-S_{\pm} \\
\label{efeA2}
N_1' =(q-4\Sigma_+)N_1 \\
\label{efeA3}
N_2' = (q+2\Sigma_++2\sqrt{3}\Sigma_-)N_2 \\
\label{efeA4}
N_3' = (q +2 \Sigma_+ -2\sqrt{3} \Sigma_-)N_3  ,
\end{eqnarray} 
where
\begin{eqnarray}
\label{defS+}
S_+= \frac{1}{6} \left[ (N_2-N_3)^2 -N_1(2N_1-N_2-N_3) \right] \\
\label{defS-}
S_-=\frac{1}{2\sqrt{3}} (N_3-N_2)(N_1-N_2-N_3)  ,
\end{eqnarray} 
together with the decoupled equation  
\begin{equation}
\label{Hprime}
H'= -(1+q)H  .
\end{equation}

Eqs. (\ref{limitKSO}) and (\ref{defSigma}) imply immediately that 
$\lim_{\tau \rightarrow -\infty} \Sigma_+ =  
\lim_{\tau \rightarrow -\infty} \Sigma_- = 0$
and it follows from Eq. (\ref{defq}) that 
$\lim_{\tau \rightarrow -\infty} q = (3\gamma -2)/2 >0$. 
By following a similar argument to that of \cite{waincoley} 
(who consider expanding cosmologies with $0 \le \gamma <2/3$), 
we may then deduce from Eqs. (\ref{efeA2})--(\ref{efeA4}) that for 
each $N_a$ there exists a parameter $\epsilon >0$ such that 
$N'_a/N_a > \epsilon$ for a sufficiently 
negative $\tau$, and hence that $\lim_{\tau \rightarrow -\infty}
N_a =0$. We conclude, therefore, that the spatially flat 
and isotropic FRW universe is the global sink 
for all ultra-stiff, orthogonal, initially contracting Bianchi 
models of type I-VIII (class A). 

\subsection{Bianchi Class B}

For the class B models, we employ the framework developed by 
Hewitt and Wainwright \cite{wainB} and define the quantities 
$\tilde{\sigma} \equiv \frac{1}{6} \tilde{\sigma}_{ab} \tilde{\sigma}^{ab}$
and $\sigma_+ \equiv \frac{1}{2} {\sigma^a}_a$, where 
$\tilde{\sigma}_{ab}$ is the trace-free part of 
the shear tensor $\sigma_{ab}$. Likewise, we define 
$\tilde{n} \equiv \frac{1}{6} \tilde{n}_{ab}\tilde{n}^{ab}$ and 
$n_+ \equiv \frac{1}{2}{n^a}_a$, where $\tilde{n}_{ab}$ 
is the trace-free part of $n_{ab}$. Since $A \ne 0$ for this class, 
there exists a constant $\tilde{h}$ such that 
$\tilde{n}=\frac{1}{3} (n_+^2 - \tilde{h} A^2)$. 
For $n_2n_3 \ne 0$, 
this defines the group parameter $h= \tilde{h}^{-1}$ for 
types ${\rm VI}_h$ $(h <0 )$ and ${\rm VII}_h$ $(h>0)$, 
respectively. (Types IV and V correspond to $\tilde{h}=0$
and the Bianchi type III is the same as the type ${\rm VI}_{-1}$).

The specific form of the Einstein field equations 
is unimportant in establishing the nature of the attractor solution for 
ultra-stiff, collapsing Bianchi class B universes. 
Indeed, it is sufficient to define Hubble normalized variables 
\cite{wainwright,wainB}
such that $\Sigma_+ = \sigma_+/H$, $\tilde{\Sigma}=\tilde{\sigma}/H$, 
$\tilde{A}=A^2/H^2$, $N_+=n_+/H$ and 
$\tilde{N}= \frac{1}{3}(N_+^2-\tilde{h} \tilde{A})$, 
where $\tilde{A} \ge 0$, $\tilde{\Sigma} \ge 0$, and $\tilde{N} 
\ge 0$. In this case, the evolution equation 
for the density parameter, $\Omega$, still has the form given by 
Eq. (\ref{Omegaevolve}), where the deceleration parameter, $q$, is
defined by  Eq. (\ref{defq}),  
but the variables $\{ K , \Sigma^2 \}$ in these expressions 
are now defined by 
\begin{eqnarray}
\label{Ktilde}
K=\tilde{N}+\tilde{A} \\
\label{tildeSigma}
\Sigma^2=\Sigma^2_+ +\tilde{\Sigma}  .
\end{eqnarray}
Inspection of Eq. (\ref{limitKSO}) therefore implies that as the 
universe approaches the big crunch, 
$\lim_{\tau \rightarrow -\infty} \Sigma_+ = 
\lim_{\tau \rightarrow -\infty} \tilde{\Sigma} =0$ and, since $\tilde{N}$ 
and $\tilde{A}$ are both non-negative quantities, it 
also follows that $\lim_{\tau \rightarrow -\infty} \tilde{N}  =
\lim_{\tau \rightarrow -\infty} \tilde{A} =0$. Moreover, 
the dimensionless curvature variables for the Bianchi class 
B are given by 
\begin{equation}
\label{curvB}
{\cal{S}}_+ = 2\tilde{N}, \qquad \tilde{\cal{S}}^{ab}
\tilde{\cal{S}}_{ab} = 24 (\tilde{A}+N_+^2)\tilde{N}  ,
\end{equation} 
from which we deduce that $\lim_{\tau \rightarrow -\infty} {\cal{S}}_+ = 
\lim_{\tau \rightarrow -\infty} \tilde{\cal{S}}^{ab}
\tilde{\cal{S}}_{ab} =0$. 

In the Bianchi class B, there is a special model corresponding to the 
type ${\rm VI}_{-1/9}$, which is exceptional in the sense that its 
isometry group does not necessarily 
admit an Abelian subgroup that acts orthogonally 
transitively. 
Nonetheless, Eqs. (\ref{Omegaevolve}) and (\ref{limitKSO}) and  
Eqs. (\ref{Ktilde})-(\ref{curvB}) remain valid for the exceptional 
type ${\rm VI}_{-1/9}$ and the above analysis 
therefore applies to all Bianchi class B models. (More precisely, 
the shear parameter (\ref{tildeSigma}) acquires additional terms due to the 
extra independent component, but it may still be expressed as a sum of 
non-negative terms that must all vanish when Eq. (\ref{limitKSO}) 
holds \cite{htw}). 
We may conclude, therefore, that the spatially flat 
and isotropic FRW universe is the global sink 
for all ultra-stiff, orthogonal, initially contracting Bianchi 
class B cosmologies. 

\subsection{Bianchi Type IX}
 
It now only remains to consider the Bianchi type IX universe.
The physical state of a Bianchi type IX model is 
parametrized by the state vector 
${\bf x} = (H, \sigma_+, \sigma_- , n_1,n_2,n_3) \in \Re^6$ as for 
other types in the class A. However, since $n_1 > 0$, $n_2 >0$, and
$n_3 >0$ for this model, the curvature parameter $K$ 
is no longer semi-positive definite and, consequently, the Hubble 
parameter may not necessarily be a monotonically 
varying function of time. This implies that alternatives 
to the Hubble-normalized variables are required for a global 
analysis. 
A set of appropriate variables that compactifies the Bianchi type IX 
phase space was introduced by 
Hewitt, Uggla and Wainwright (see section 8.5.2 of \cite{wainellis}). 
In this approach, physical quantities are normalized 
in terms of a function, $D$, such that  
\begin{equation}
\label{normvars}
(\bar{H} , \bar{\Sigma}_{\pm} , \bar{N}_a , \bar{\Omega} ) \equiv  
\left( \frac{H}{D} , \frac{\sigma_{\pm}}{D} , \frac{n_a}{D} , 
\frac{\rho}{3D^2} \right)  ,
\end{equation}
where 
\begin{equation}
\label{defD}
D \equiv \sqrt{H^2+\frac{1}{4}(n_1n_2n_3)^{2/3}}  .
\end{equation}
In particular, Eq. (\ref{defD}) implies that $\bar{H}$ is bounded 
by the constraint equation
\begin{equation}
\label{defbarH}
\bar{H}^2+ \frac{1}{4} ( \bar{N}_1\bar{N}_2\bar{N}_3 )^{2/3} =1  .
\end{equation}

A new time variable, $\bar{\tau}$, is also defined:
\begin{equation}
\label{bartau}
\frac{dt}{d\bar{\tau}} = \frac{1}{D}
\end{equation}
and Eqs. (\ref{efeA2})-(\ref{efeA4}), (\ref{Hprime}) and (\ref{defD}) 
then imply that 
the evolution equation for $D$ takes the form
\begin{equation}
\label{Devolve}
D^* = - (1+\bar{q})\bar{H}D  ,
\end{equation}
where a star
denotes differentiation with respect to $\bar{\tau}$ and  
$\bar{q} \equiv \bar{H}^2q$. 
It can be further shown that the Einstein field equations 
take the form \cite{wainellis}
\begin{eqnarray}
\label{IXa}
\bar{H}^*=-(1-\bar{H}^2)\bar{q} \\
\label{IXb}
\bar{\Sigma}^*_{\pm} = -(2-\bar{q}) \bar{H} \bar{\Sigma}_{\pm}
-\bar{S}_{\pm} \\
\label{IXc}
\bar{N}^*_1 = (\bar{H}\bar{q} -4\bar{\Sigma}_+)\bar{N}_1 \\
\label{IXd}
\bar{N}^*_2 = (\bar{H}\bar{q} +2\bar{\Sigma}_+ +2\sqrt{3}
\bar{\Sigma}_- ) \bar{N}_2 \\
\label{IXe}
\bar{N}^*_3 =(\bar{H} \bar{q} +2\bar{\Sigma}_+ -2 \sqrt{3}
\bar{\Sigma}_- ) \bar{N}_3  ,
\end{eqnarray}
where $\bar{S}_{\pm}$ are defined by Eqs. (\ref{defS+})
and (\ref{defS-}), respectively, with $N_a$ replaced by 
$\bar{N}_a$. 

Moreover, it follows from Eqs. (\ref{defq}) and (\ref{defbarH}) that 
\begin{equation}
\label{defbarq}
\bar{q} = \frac{1}{2} (3\gamma -2) (1-\bar{V} ) +\frac{3}{2}
(2-\gamma ) \bar{\Sigma}^2  ,
\end{equation}
where 
\begin{equation}
\label{defbarSigma}
\bar{\Sigma}^2 = \bar{\Sigma}^2_+ +\bar{\Sigma}_-^2
\end{equation}
and 
\begin{eqnarray}
\label{defbarV}
\bar{V} \equiv \frac{1}{12} \left[ \bar{N}_1^2 +\bar{N}_2^2 
+\bar{N}_3^2 -2\bar{N}_1\bar{N}_2 -2 \bar{N}_2\bar{N}_3 
-2\bar{N}_1\bar{N}_3 \right. 
\nonumber 
\\ 
\left. +3(\bar{N}_1\bar{N}_2\bar{N}_3 )^{2/3} \right]  .
\end{eqnarray}
The definition (\ref{defbarV}) implies that $\bar{V} \ge 0$ 
and substitution of Eq. (\ref{defK}) yields $\bar{V}=\bar{H}^2 [K
+(N_1N_2N_3)^{2/3}/4]$. Thus, the Friedmann constraint equation 
(\ref{genfried}) may be expressed in the form: 
\begin{equation}
\label{bargenfried}
\bar{\Sigma}^2+\bar{V} +\bar{\Omega} =1  ,
\end{equation}
whereas substitution of Eq. (\ref{bargenfried}) into 
Eq. (\ref{defbarq}) implies that 
\begin{equation}
\label{qpos}
\bar{q} =2\bar{\Sigma}^2 +\frac{1}{2} (3\gamma -2) \bar{\Omega}   .
\end{equation}

Finally, we may derive an evolution equation for the density 
parameter, $\bar{\Omega} = \rho/(3D^2)$, by differentiating 
with respect to $\bar{\tau}$, and substituting in the 
fluid equation (\ref{fluid}) and the evolution equation (\ref{Devolve}). 
We find that 
\begin{equation}
\label{hatbarOmega}
\bar{\Omega}^* = \bar{\Omega} \bar{H} 
\left[ -(3\gamma -2) \bar{V} +3(2-\gamma ) \bar{\Sigma}^2 \right]  .
\end{equation}

We may now deduce from Eq. (\ref{qpos}) that when $(\gamma -2)$ 
is positive--definite, $\bar{q} \ge 0$ with equality iff 
$\bar{\Omega} = \bar{\Sigma}^2 =0$. Furthermore, 
since Eq. (\ref{defbarH}) implies that $\bar{H}$ is bounded,
$-1\le \bar{H} \le 1$, 
it follows from Eq. (\ref{IXa}) that $\bar{H}^* \le 0$ and, consequently, 
that $\bar{H}$ is a monotone decreasing function. On the other hand, 
$\bar{q} >0$ for a non--vacuum orbit $(\bar{\Omega} >0)$, and 
this implies that $\lim_{\bar{\tau} \rightarrow - \infty} 
\bar{H} =1$ and $\lim_{\bar{\tau} \rightarrow + \infty} 
\bar{H} = -1$. In other words, an initially 
expanding model will eventually undergo a recollapse (when 
$\bar{H}$ passes through zero). Once the recollapse occurs,  
Eq. (\ref{hatbarOmega}) implies that $\bar{\Omega}^* \ge 0$, 
with equality iff $\bar{V} =\bar{\Sigma}^2=0$. However, 
since $\bar{\Omega} \le 1$ is bounded due to Eq. 
(\ref{bargenfried}), we deduce that  
$\lim_{\bar{\tau} \rightarrow +\infty} \bar{\Omega}^*
=0$ and, hence, that
\begin{equation}
\lim_{\bar{\tau} \rightarrow +\infty} \bar{\Omega} =1 , \quad 
\lim_{\bar{\tau} \rightarrow +\infty} \bar{V}=0 , \quad 
\lim_{\bar{\tau} \rightarrow +\infty} \bar{\Sigma}^2 =0 .
\end{equation}

Thus, Eq. (\ref{qpos}) implies that 
$\lim_{\bar{\tau} \rightarrow +\infty} \bar{q} = 
(3\gamma -2)/2$, from which it follows via 
Eqs. (\ref{IXc})--(\ref{IXe}) that 
for a sufficiently large $\bar{\tau}$, there exists an 
$\epsilon >0$ such that $d \ln \bar{N}_a/d\bar{\tau} <-\epsilon$
and therefore that $\lim_{\bar{\tau} \rightarrow +\infty} \bar{N}_a =0$.
Consequently, a Bianchi type IX cosmology sourced by an ultra-stiff fluid 
isotropizes in the same way as the other Bianchi types as it 
approaches the big crunch at $\bar{\tau} \rightarrow \infty$. 
In particular, this implies that there is no chaotic (oscillatory)
behaviour in the vicinity of the singularity. 

Since we have now covered all possible Bianchi types,
we may summarize the above analysis in the 
form of a cosmic no hair theorem: 
{\em all initially contracting, spatially homogeneous, orthogonal 
Bianchi type I-VIII cosmologies and all Bianchi type IX universes that are
sourced by an ultra-stiff fluid with an equation of state such that 
$(\gamma -2 )$ is positive-definite collapse into an isotropic singularity, 
where the sink is the spatially flat and isotropic FRW universe.} 
For the class of models where the equation of state asymptotes to a constant 
value $\gamma >2$ on the approach to the singularity, the FRW cosmology 
is a self-similar power-law solution, where the scale factor varies as  
$a \propto (-t)^{2/3\gamma}$.  
 
In the following Section, we employ this theorem to determine 
the nature of a collapsing Bianchi type IX universe  with a 
matter source consisting of a minimally coupled 
scalar field self-interacting through a negative-definite exponential 
potential. 

\section{Isotropization of Collapsing Bianchi type IX 
Scalar Field Cosmology}

In this Section, we consider an action of the form 
\begin{equation}
S=\int d^4x \sqrt{-g} \left[ R- \left( \nabla \varphi \right)^2 
- V(\varphi ) \right] , \qquad V=V_0\exp (- \lambda \varphi )  ,
\end{equation}
where $V$ represents the potential 
of the scalar field, $\varphi$, and 
$V_0 <0$ and $\lambda >0$ are constants. 
We assume the field is orthogonal, i.e., that it is 
constant on the surfaces of homogeneity, $\varphi =\varphi (t)$, 
so that its energy density is given by $\rho = \frac{1}{2}
\dot{\varphi}^2 +V$. The effective equation of state of the field 
is defined by
\begin{equation}
\gamma = \frac{2 \dot{\varphi}^2}{\dot{\varphi}^2 +2V}
\end{equation}
and is therefore bounded such that $\gamma \ge 2$.

The presence of a negative potential energy implies that not all the 
terms on the right-hand side of the Friedmann equation (\ref{Fried}) 
will be semi-positive definite. Thus, a given Bianchi model 
may undergo a recollapse and, consequently, the Hubble normalized 
variables do not lead to a global compact phase space. Nonetheless, 
the theorem developed in the previous Section can be employed to 
determine the nature of collapsing, homogeneous scalar field cosmologies 
in the vicinity of the big crunch singularity. 

It is known that the collapsing Bianchi I model 
is stable to curvature and anisotropy perturbations 
near the singularity \cite{wands}. 
In this Section, we will consider the Bianchi type IX universe. In this case,  
variables parametrizing the evolution 
of the scalar field can be defined such that \cite{hch}
\begin{equation}
\label{defPsi}
\bar{\Psi} \equiv \frac{\dot{\varphi}}{\sqrt{6}D} , \qquad 
\bar{\Theta} \equiv \frac{\sqrt{-V}}{\sqrt{3}D} .
\end{equation}
The evolution equations for these variables are then 
determined by the scalar field equation, $\dot{\rho} = - 3H\dot{\phi}^2$, 
which itself follows as a consequence of energy-momentum conservation. 
We find that 
\begin{eqnarray}
\label{Psievolve}
\bar{\Psi}^* = (\bar{q}-2) \bar{H} \bar{\Psi} - \frac{\sqrt{6}\lambda}{2} 
\bar{\Theta}^2 \\
\label{Thetaevolve}
\bar{\Theta}^*= \left[ (1+\bar{q}) \bar{H} - \frac{\sqrt{6}\lambda}{2}
\bar{\Psi} \right] \bar{\Theta} .
\end{eqnarray}

It also proves convenient to 
define a new bounded variable \cite{hch}: 
\begin{equation}
\label{defd}
d \equiv \frac{D}{D+1}  ,
\end{equation}
where $0 \le d \le 1$. This evolves such that 
\begin{equation}
\label{devolve}
d^* = (1+\bar{q} ) \bar{H} d(d-1)
\end{equation}
and the big crunch singularity, $\bar{H} \rightarrow -1$, 
corresponds to $D \rightarrow  \infty$ 
$(d\rightarrow 1)$. 

The physical state of a Bianchi type IX scalar field cosmology  
is therefore given by the vector ${\bf x} = (d, \bar{H}, \bar{\Sigma}_+,
\bar{\Sigma}_-,\bar{N}_1,\bar{N}_2,\bar{N}_3, \bar{\Psi},\bar{\Theta})$
and the corresponding autonomous set of ODEs are 
Eqs. (\ref{IXa})-(\ref{IXe}) and Eqs. (\ref{Psievolve}), (\ref{Thetaevolve})
and (\ref{devolve}), 
subject to the constraint equations (\ref{defbarH}) and (\ref{bargenfried}).  
The density parameter is $\bar{\Omega} = 
\bar{\Psi}^2-\bar{\Theta}^2$ and evolves according to 
Eq. (\ref{hatbarOmega}). 

We will assume that 
the universe is undergoing a collapse, $\bar{H}<0$, and that 
the field has positive energy density, $\bar{\Omega}>0$. It is 
anticipated that this should represent the behaviour of generic solutions 
once the recollapse has set in. Since $\bar{q}>0$, Eq. (\ref{IXa}) 
implies that $\bar{H}^*<0$ and, consequently, Eq. (\ref{hatbarOmega}) 
implies that $\bar{\Omega}$ continues to grow monotonically. 
Thus, the energy density of the field remains positive, 
$\dot{\phi}^2 > 2|V|$, and the sign of $\bar{\Psi}$ is fixed 
on the approach to the big crunch. 
Moreover, since $\bar{\Omega}>0$ and $\bar{q}>0$, the variables 
$\{ \bar{\Sigma}^2_{\pm},\bar{V} , \bar{\Omega} \} \le 1$ (due to 
Eq. (\ref{bargenfried})) and 
$\lim_{\bar{\tau} \rightarrow \infty} \bar{H}  =-1$.  

The cosmic no hair theorem of Section II will apply for this 
collapsing Bianchi type IX scalar field cosmology
if it can be shown that $\gamma >2$ at the singularity. 
In principle, however, a scalar field may become dominated 
by its kinetic energy during the 
collapse, in which case $\gamma \rightarrow 2$ and $\bar{\Theta} 
\rightarrow 0$. 
This implies that the shear may not necessarily become sub-dominant. 
In order to establish whether this possibility arises, 
we will first show that if 
$\lim_{\bar{\tau} \rightarrow \infty} \gamma =2$, 
the contracting Bianchi IX
model asymptotes in the vicinity of the big 
crunch toward the spatially 
flat type I background, where $\bar{N}_1=\bar{N}_2=\bar{N}_3=0$. 

Firstly, an argument similar to that presented after Eq. 
(\ref{hatbarOmega}) implies that  
$\lim_{\bar{\tau} \rightarrow \infty} \bar{V} =0$. 
However, for an equilibrium point with $\gamma =2$, 
Eq. (\ref{defbarq}) implies that $\bar{q} =2$. It then follows
from Eq. (\ref{IXb}) that this equilibrium point will also have
$\bar{S}_{\pm} =0$. The definition (\ref{defS-}) 
then requires that $\bar{N}_3 =\bar{N}_2$ 
or $\bar{N}_1 =\bar{N}_2+\bar{N}_3$. If the latter 
condition applies, the requirements that $\bar{V} =\bar{S}_+=0$ then imply 
that $\bar{N}_2 =0$ (assuming without loss 
of generality that $\bar{N}_3>\bar{N}_2>\bar{N}_1$). 
On the other hand, if $\bar{N}_1 =\bar{N}_3 \ne 0$, 
Eqs. (\ref{IXc}) and (\ref{IXe}) imply that $\bar{\Sigma}_+ = -1/2$
and $\bar{\Sigma}_-= -\sqrt{3}/2$. In other words, 
the equilibrium point would correspond to a vacuum model,
where $\bar{\Psi} =\bar{\Theta}=0$ due to the constraint (\ref{bargenfried}). 
This is inconsistent with the property that $\bar{\Omega}^* \ge 0$ as 
$\bar{\tau}\rightarrow +\infty$. 

Imposing the alternative condition $\bar{N}_2 = \bar{N}_3$
reduces the constraint $\bar{S}_+=0$ to $\bar{N}_1(\bar{N}_1-\bar{N}_2 ) =0$. 
Hence, either $\bar{N}_1=\bar{N}_2=\bar{N}_3$ or 
$\bar{N}_1=0$. However, if $\bar{N}_2= \bar{N}_3 \ne 0$, 
Eqs. (\ref{IXd}) and (\ref{IXe}) imply that
$\bar{\Sigma}_- =0$ and $\bar{\Sigma}_+ =1$, and 
once more this would correspond to a vacuum model, which is 
a contradiction. 

Hence, the only equilibrium point (with $\bar{H}<0$) that 
corresponds to a non-vacuum cosmology  
with $\gamma =2$ lies in the Bianchi type I 
invariant set and is given by  
\begin{eqnarray}
\label{eqmpoint}
d =1 , \quad \bar{H}=-1, \quad \bar{N}_1=\bar{N}_2=\bar{N}_3=0, 
\nonumber \\
\bar{q}=2, \quad \bar{\Theta} =0, \quad 
\bar{\Sigma}_+^2+ \bar{\Sigma}_-^2 +\bar{\Psi}^2=1  .
\end{eqnarray}  
A standard perturbation analysis reveals that 
the eigenvalues associated with this point are
\begin{eqnarray}
\label{eigenvalues}
-3, \quad -4 , \quad 0, \quad 0, \quad -2-4\bar{\Sigma}_+,
\quad -4, 
\nonumber 
\\ 
\quad -2 +2\bar{\Sigma}_+  \pm 2\sqrt{3}  \bar{\Sigma}_- , 
 \quad -3 -\frac{\sqrt{6}\lambda}{2} \bar{\Psi}  .
\end{eqnarray}
The two vanishing eigenvalues imply that this is a two-dimensional 
set of equilibrium points. 
It follows that the stability of this set of points is sensitive to the 
kinetic energy of the scalar field. If $\bar{\Psi} < -\sqrt{6}/\lambda$, 
it represents a saddle, otherwise it may  be a source if 
$\bar{\Sigma}_+> 1/2$ and $\bar{\Sigma}_-$ is sufficiently 
small. 

As there are no other equilibrium points, this  
implies in effect that if the scalar field is rolling to 
large negative values down its potential at a sufficiently fast rate,
$\bar{\Psi} < -\sqrt{6}/\lambda$, 
its equation of state will satisfy 
$\lim_{\bar{\tau} \rightarrow \infty} \gamma >2$. We may therefore apply 
the cosmic no hair theorem of the previous 
Section in this region of parameter space. 
In other words, the stable attractor solution  
is the spatially flat and isotropic FRW cosmology
corresponding to the equilibrium point  
$\bar{\Psi} = -\lambda / \sqrt{6}$, 
$\bar{\Theta} = [(\lambda^2 /6) -1]^{1/2}$ 
and $\bar{q}=(\lambda^2/2) -1$. This `scaling' solution  
exists only for $\lambda > \sqrt{6}$.
In conclusion, therefore, a contracting Bianchi IX cosmology
sourced by an orthogonal scalar field rolling down a negative exponential 
potential with $\bar{\Psi} < -\sqrt{6}/\lambda$ and 
$\lambda >\sqrt{6}$ collapses into an isotropic singularity, 
where the sink is the self-similar, 
spatially flat and isotropic FRW universe
with scale factor $a \propto (-t)^{2/\lambda^2}$. 

In the next Section we consider such a scalar field model 
generated from a compactified  higher-dimensional cosmology
inspired by ${\rm M}^*$-theory. 

\section{Collapsing M$^*$-Cosmology}

The ${\rm M}^*$-theory of Hull \cite{hull} is a version of 
M-theory in an eleven-dimensional spacetime of signature 
(9+2). Compactification of this theory on a timelike circle 
gives rise to a string theory -- termed the type ${\rm IIA}^*$-theory -- 
with a supergravity limit that is similar to the conventional 
type IIA string but with a Ramond-Ramond (RR) sector 
containing form-fields whose kinetic terms have the `wrong' sign. In this 
sense, ${\rm M}^*$-theory may be interpreted as the 
strongly-coupled limit of the type ${\rm IIA}^*$ theory. Its low-energy 
limit is a supergravity theory with a bosonic sector given by 
\begin{equation}
\label{M*action}
S_{\rm M^*} =\int d^{11} x \sqrt{|g|} \left[ R + \frac{G_4^2}{48} \right]
-\frac{1}{12} \int C_3 \wedge G_4 \wedge G_4  ,
\end{equation}
where $G_4$ represents the field strength of the three--form potential 
$C_3$. The sign of the kinetic term for $G_4$ is determined by 
supersymmetry. 

Leaving aside issues associated with the conceptual 
problems of extra timelike dimensions (see \cite{hull} for a full 
discussion of such questions within the 
context of ${\rm M}^*$-theory), we may 
consider the Kaluza-Klein compactification 
of the action (\ref{M*action}) on a timelike circle $S^1$. 
For the case where only the 
RR four-form field strength is non-trivial, 
the ten-dimensional action for the
truncated type ${\rm IIA}^*$-theory is given by 
\begin{equation}
\label{IIAaction}
S_{\rm IIA^*} = \int d^{10} x \, \sqrt{-g_s} \left[ 
e^{-\phi_{10}} \left( R_s +\left( \nabla \phi_{10} \right)^2 \right) 
+\frac{G_4^2}{48} \right]  ,
\end{equation}
where the ten-dimensional dilaton field, $\phi_{10}$, is related to the 
radius of the eleventh dimension, $e^{r_{11}}$, by $\phi_{10} = 3r_{11}$
and we have performed a conformal transformation: 
\begin{equation}
\label{11dconftrans}
g^{(s)}_{AB} = \Omega^2 g_{AB}, \qquad \Omega^2 \equiv e^{r_{11}}
\end{equation}
to the string-frame metric, $g^{(s)}_{AB}$. 

We will now dimensionally reduce the theory (\ref{IIAaction}) 
to four dimensions on a six-torus, $T^6$, where the only dynamical 
degree of freedom in the internal dimensions is taken to 
be the breathing mode, $\beta$. In other words,  
we assume that the string-frame metric (\ref{11dconftrans}) 
is given by 
\begin{equation}
\label{stringmetric}
ds^2_s=g^{(s)}_{\mu\nu} (x) dx^{\mu}dx^{\nu} +e^{2 \beta (x)} 
dy^2 ,
\end{equation}
where $dy^2 = \delta_{ij}dy^idy^j$. 
We will also Hodge dualize the four-form field strength in ten dimensions 
to a six-form, $F_6$, and assume that the only non-zero components 
of this six-form live on the internal dimensions, 
i.e., we assume an {\em ansatz} $F_6 = m \epsilon_6$, where $\epsilon_6$ 
is the volume-form of $T^6$ and $m$ is a constant. 
Thus, the effective four--dimensional effective action takes the form
\begin{equation}
\label{4dstringaction}
S= \int d^4 x \sqrt{-g_s} \left[ e^{-\phi_4} \left( R_s +\left( \nabla 
\phi_4 \right)^2 - 6\left( \nabla \beta \right)^2 \right) 
-\frac{1}{2} m^2 e^{-6\beta} \right] ,
\end{equation}
where $\phi_4 \equiv \phi_{10} -6\beta$ represents the four-dimensional 
dilaton field. 

Eq. (\ref{4dstringaction}) may be expressed in 
the Einstein-Hilbert form by performing the conformal transformation
\begin{equation}
\label{4dconftrans}
g^{(e)}_{\mu\nu} = e^{-\phi_4} g^{(s)}_{\mu\nu}
\end{equation}
and field redefinitions $\tilde{\phi}_4 = \phi_4/\sqrt{2}$ and 
$\tilde{\beta} = \sqrt{6} \beta$. It follows 
that the conformally transformed action is given by  
\begin{equation}
\label{4deinsteinaction}
S= \int d^4x \, \sqrt{-g_e} \left[ R_e -\left( \nabla \tilde{\phi}_4 \right)^2
- \left( \nabla \tilde{\beta} \right)^2 + m^2
e^{\sqrt{8} \tilde{\phi}_4 - \sqrt{6} \tilde{\beta}  }
\right]  .
\end{equation}
Finally, defining a new pair of scalar fields:   
\begin{eqnarray}
\label{fieldredefine}
\chi \equiv - \frac{1}{\sqrt{14}} \left( \sqrt{8} \tilde{\phi}_4 - \sqrt{6} 
\tilde{\beta} \right)
\nonumber \\
\xi \equiv \frac{1}{\sqrt{14}} \left( \sqrt{6} \tilde{\phi}_4 + 
\sqrt{8} \tilde{\beta} \right)
\end{eqnarray}
implies that the action (\ref{4deinsteinaction}) is equivalent to 
\begin{equation}
\label{rescaleaction}
S= \int d^4x \, \sqrt{-g_e} \left[ R_e -\left( \nabla \chi \right)^2
- \left( \nabla \xi \right)^2  + m^2 e^{- \sqrt{14} \chi}
\right]  .
\end{equation}

Thus, the potential for the $\chi$-field is negative and 
sufficiently steep ($\lambda = \sqrt{14}$) for the results 
of Section III to apply. 
The scalar degree of freedom $\xi$ will behave as a stiff perfect fluid, but 
the presence of this field does not affect the 
conclusions of the previous section, since it is decoupled 
from the other matter degree of freedom. It therefore becomes 
dynamically negligible near to the singularity in the case 
where the collapsing Bianchi type IX cosmology asymptotes 
to the FRW solution. In this case, the attractor is the power law solution 
$a_e \propto (-t_e)^{1/7}$, $\chi = (2/\sqrt{14} ) \ln (-t_e)$, 
$\xi =0$. Transforming back into the four-dimensional 
string frame via Eq. (\ref{4dconftrans}) then implies that 
$a_s \propto (-t_s)^{-1/5}$, $e^{\phi_4} \propto 
(-t_s)^{-4/5}$ and $e^{\beta} \propto (-t_s)^{1/5}$. 
The ten-dimensional string-frame metric (\ref{stringmetric})
is therefore given by 
\begin{equation}
\label{IIA*solution}
ds_s^2 = T^{-1/2}_s dx_3^2 +T^{1/2}_s \left[ -dT^2_s +dy^2_6 \right]  ,
\end{equation}
where we have defined $T_s \equiv (-t_s)^{4/5}$ and rescaled the coordinates 
where appropriate. 

The scale factors in the eleven-dimensional frame, 
which we may denote by $\{ \hat{a} , e^{\hat{\beta}} , e^{r_{11}} \}$
are related to the corresponding ten-dimensional quantities  
by Eq. (\ref{11dconftrans}) and take the form $\hat{a} =a e^{-r_{11} /2}$ and 
$\hat{\beta} =\beta -r_{11} /2$, where the eleven-dimensional proper time 
is given by $\eta \equiv \int dt_s \, \exp (-r_{11} /2 )$. 
Hence, the eleven-dimensional line element can be expressed as 
\begin{equation}
\label{M*solution}
ds_{11}^2  =  \hat{\eta}^{-2/3} dx^2_3 + \hat{\eta}^{1/3}
\left[ -d\hat{\eta}^2 -dt_{11}^2 +dy^2_6 \right]  ,
\end{equation}
where $\hat{\eta} \equiv (-\eta)^{6/7}$, the 
coordinate of the eleventh (timelike) dimension is denoted by 
$t_{11}$ and, for simplicity, we have absorbed any 
constants of proportionality by a suitable rescaling of the 
spacetime coordinates.  

A background of the form (\ref{IIA*solution}) 
can be interpreted as the cosmological analogue of a domain wall 
spacetime \cite{hullthree}. More specifically, we may write the metric 
in the form $ds_{11}^2 = F^{-2/3} dx^2_3 +F^{1/3}
[ -d\hat{\eta}^2 -dt_{11}^2 +dy^2_6 ]$, where $\{ \hat{\eta}, t_{11}, 
y\}$ represent the transverse coordinates. Since $F$ is a linear function 
of $\hat{\eta}$, it may be viewed as a harmonic function 
on the transverse space, thereby allowing 
Eq. (\ref{M*solution}) to be interpreted as a membrane-type solution with a 
three-dimensional Euclidean world-volume \cite{hullkhuri,hullthree}. 
In particular, it is known that solutions to ${\rm M}^*$-theory 
of this form preserve 16 of the 32 supersymmetries of the 
theory \cite{hullthree}. It is interesting that such a supersymmetric, 
higher-dimensional cosmological background is 
selected from a dynamical 
point of view when analyzed in terms of an 
effective four-dimensional, anisotropic cosmology.

\section{Conclusion}

To summarize, it has been shown that 
any orthogonal, collapsing Bianchi cosmology with a matter sector 
comprised of an ultra-stiff fluid with 
an arbitrary and varying equation of state (subject to the condition 
that $\gamma >2$ at all times)
approaches spatial isotropy and flatness in the neighbourhood of 
the big crunch singularity. 
Of particular interest is the behaviour of the general 
Bianchi type IX universe. Since the 
singularity is isotropic, such models do not exhibit chaotic-type (Mixmaster)
behaviour near to the big crunch, as is the case when $\gamma <2$
(see, e.g., \cite{ringstrom} and references therein). A related 
effect has been found in a class of braneworld models with a constant 
equation of state parameter $\gamma >1$, 
where the effective Friedmann equation depends quadratically 
on the energy density \cite{coley}. This can be understood since 
the fluid in such a model behaves at small spatial volumes as if 
it was ultra-stiff \cite{lidseycopeland}. 

The only spatially homogeneous cosmology that we have not considered 
is the Kantowski-Sachs model, where the $G_3$ group of isometries 
acts multiply-transitively. The line element for this cosmology may be 
written in the form
\begin{eqnarray}
ds^2 =-dt^2 +D_1^2dx^2 +D_2^2 d\Omega_2^2 \nonumber \\
\label{KSmetric}
D_1(t) = e^{\beta_0 (t) -2\beta_+(t) } , \qquad 
D_2(t) = e^{\beta_0 (t) +\beta_+ (t)}
\end{eqnarray}
where $d\Omega_2^2$ is the metric on the two-sphere. Since this spacetime 
has positive spatial curvature, an extension of our cosmic 
no hair theorem to include this case would 
require a set of variables that compactifies the 
phase space in an analogous way to the variables (\ref{normvars}) 
and (\ref{defD}) for the Bianchi type IX model. However, 
for $\gamma >2$, a suitably normalized variable for the energy density that 
increases monotonically during a collapsing phase has yet 
to be identified. 

Nonetheless, we may gain insight into the asymptotic nature of 
an ultra-stiff Kantowski-Sachs universe by assuming $\gamma = {\rm constant}$. 
Following \cite{coleybook}, 
we define $B_{1,2} \equiv D_{1,2}^{-1}$. The Friedmann
equation then takes the form
\begin{equation}
\label{KSFriedmann}
3H^2 = \rho +\frac{1}{3} \sigma_+^2 -B^2_2 
\end{equation}
where $H = \dot{\beta}_0$ and $\sigma_+ = 3\dot{\beta}_+$. 
It follows that $D \equiv [9H^2+3B_2^2 ]^{1/2}$ is a dominant 
variable (assuming $\rho \ge 0$), and this implies that 
compact variables $Q_0 \equiv 3H/D$, $Q_+\equiv \sigma_+/D$ 
and $\Omega \equiv 3\rho/D^2= 1-Q_+^2$ may be introduced, together with
a new time variable $\tau = \int dt D/3$. 
The curvature variable is $\tilde{K} = B_2^2/(3H^2) = (1-Q_0^2)/Q^2_0$ 
from which we deduce that $0\le \{ Q_0 ,Q_+, \Omega \} \le 1$. 

It can then be shown \cite{coleybook} 
that the field equations reduce to the set of evolution 
equations
\begin{eqnarray}
\label{KSequation1}
\frac{dQ_0}{d\tau} =- \left( 1-Q_0^2 \right) F
\\
\label{KSequation2}
\frac{dQ_+}{d\tau} = -1+\left( Q_0-Q_+ \right)^2 +Q_0Q_+F
\end{eqnarray}
where 
\begin{equation}
\label{defF}
F \equiv \frac{3\gamma}{2} -1 -Q_0Q_+ +\frac{3}{2} 
(2-\gamma) Q_+^2
\end{equation}

Eq. (\ref{KSequation1}) then implies that 
all equilibrium points are located at $Q_0^2=1$ or $F=0$. Let 
us first assume $Q_0^2\ne 1$ and $F=0$. It follows from 
Eqs. (\ref{KSequation2}) and (\ref{defF}) that 
$Q_+ = \pm (3\gamma -2)/(4-3\gamma)$, but these points are unphysical 
for $\gamma >2$. Since we are interested in collapsing universes, 
we therefore consider the case $Q_0=-1$. In this case, 
there are three equilibrium points, where $Q_+=0, \pm 1$. 
A stability analysis indicates that the points $Q_+=\pm 1$ are saddles 
and the point $(Q_0,Q_+)=(-1,0)$ is a sink. This latter point 
corresponds to the isotropic, spatially flat FRW cosmology. Hence, 
in the case of a constant, ultra-stiff equation of state, 
the unique attractor for a collapsing Kantowski-Sachs 
cosmology is the isotropic and spatially flat universe. 
This provides strong support for the conjecture that 
this background represents 
the attractor for a general ultra-stiff equation of state. 
It also represents strong motivation 
for searching for an appropriate set of variables 
along the lines of Eqs. (\ref{normvars})-(\ref{defD}).

Our no hair results are also 
of relevance to the nature of spacetime singularities in 
more general inhomogeneous backgrounds, since according to 
the Belinskii, Khalatnikov and Lifshitz (BKL) conjecture \cite{bkl}, 
each spatial point of an inhomogeneous universe evolves effectively 
as a Bianchi model on the approach to the singularity. 
In principle, the results derived above can be employed to gain further 
insight into the nature of inhomogeneous cosmologies 
\cite{coley,dunsby,coleylim}. Indeed, similar issues to those considered 
in this paper have been addressed by Coley and Lim \cite{coleylim}, 
who performed an asymptotic analysis of stiff and ultra-stiff 
Abelian $G_2$ and general $G_0$ spatially inhomogeneous cosmologies,
and found that the spatially flat FRW solution is locally 
stable in the past for $\gamma >2$. Our results differ 
in that it was assumed that $\gamma$ is a constant in \cite{coleylim}, 
whereas we have assumed an arbitrary (differentiable) equation of state 
$\gamma = \gamma (\rho) >2$. 
Furthermore, our no hair result for the spatially homogeneous 
Bianchi models is a global result, in the sense that the flat 
FRW solution is the unique attractor at the big crunch. 

Finally, the work of \cite{coleylim} is restricted to that 
of a perfect fluid matter source. We extended our analysis 
to the case of a collapsing, spatially
homogeneous Bianchi type IX model containing 
a scalar field that self-interacts through 
a negative potential. In this case, the effective equation of state 
varies but satisfies $\gamma \ge 2$, with 
equality corresponding to a vanishing potential. Our 
cosmic no hair theorem implies that 
the problem of demonstrating that such models 
isotropize as they collapse has been reduced to verifying that 
the potential remains dynamically significant near to the 
singularity. We have shown that this is possible, for example,  
in the case of a steep, negative exponential potential. 
Exponential potentials are of particular interest since they are common 
in cosmological models inspired by string/M-theory. One 
such theory that leads naturally to negative potentials is ${\rm M}^*$-theory, 
where the potential is generated by non-trivial flux of the form fields
\cite{hull}. We considered a truncated, dimensionally reduced effective 
action for this theory and found that when the isotropic attractor 
for a collapsing Bianchi type IX cosmology is oxidized back to 
eleven dimensions, the metric is a supersymmetric solution of 
the higher-dimensional 
theory. This is of interest since supersymmetric backgrounds 
are non-perturbatively exact solutions and therefore provide a 
valuable framework for investigating the cosmic dynamics 
in the high curvature regime near to the singularity.  
We expect that other attractor solutions for more 
general dimensional reduction schemes will also exhibit 
similar supersymmetric properties when interpreted in a ten-
or eleven-dimensional context.


\begin{thebibliography}{99}

\bibitem{ven}
M. Gasperini and G. Veneziano,
Phys. Rep. {\bf 373}, 1 (2003), arXiv:hep-th/0207130. 

\bibitem{lidsey}
J. E. Lidsey, D. Wands and E. J. Copeland, 
Phys. Rep. {\bf 337}, 343 (2000), arXiv:hep-th/9909061. 

\bibitem{cyclic}
J. Khoury, B. A. Ovrut, P. J. Steinhardt, and N. Turok, 
Phys. Rev. D {\bf 64}, 123522 (2001), arXiv:hep-th/0103239; 
P. J. Steinhardt and N. Turok, 
Phys. Rev. D {\bf 65}, 126003 (2002), arXiv:hep-th/0111098; 
J. Khoury, P. J. Steinhardt, and N. Turok, Phys. Rev. Lett. 
{\bf 92}, 031302 (2004), arXiv:hep-th/0307132. 

\bibitem{ewst}
J. K. Erickson, D. H. Wesley, P. J. Steinhardt, and N. Turok, 
Phys. Rev. D {\bf 69}, 063514 (2004), arXiv:hep-th/0312009. 

\bibitem{hull}
C. M. Hull, JHEP {\bf 9807}, 021 (1998), arXiv:hep-th/9806146;
C. M. Hull, JHEP {\bf 9811}, 017 (1998), arXiv:hep-th/9807127.

\bibitem{witten}
E. Witten, Nucl. Phys. {\bf B443}, 85 (1995), arXiv:hep-th/9503124.

\bibitem{ellismac}
G. F. R. Ellis and M. A. H. MacCallum, Commun. Math. Phys. {\bf 12}, 108 
(1969). 

\bibitem{wainA}
J. Wainwright and L. Hsu, Class. Quant. Grav. {\bf 6}, 1409 (1989). 

\bibitem{wainwright}
J. Wainwright, in {\em Relativity Today}, Proceedings of Second 
Hungarian Relativity Workshop, ed. Z. Perjes (World Scientific, 1988). 

\bibitem{wainB}
C. G. Hewitt and J. Wainwright, Class. Quant. Grav. {\bf 10}, 99 (1993).

\bibitem{wainellis}
J. Wainwright and G. F. R. Ellis (eds.) {\em Dynamical Systems in Cosmology}
(Cambridge: Cambridge University Press, 1997). 

\bibitem{waincoley}
A. A. Coley and J. Wainwright, Class. Quant. Grav. {\bf 9},
651 (1992). 

\bibitem{htw}
C. G. Hewitt, J. T. Horwood and J. Wainwright,
Class. Quant. Grav. {\bf 20}, 1743 (2003), arXiv:gr-qc/0211071.

\bibitem{wands}
I. P. C. Heard and D. Wands, 
Class. Quant. Grav. {\bf 19}, 5435 (2002), arXiv:gr-qc/0206085.

\bibitem{hch}
R. J. van den Hoogen, A. A. Coley and Y. He,
Phys. Rev. D {\bf 68}, 023502 (2003), arXiv:gr-qc/0212094. 

\bibitem{hullkhuri}
C. M. Hull and R. R. Khuri, 
Nucl. Phys. {\bf B536}, 219 (1998), arXiv:hep-th/9808069.
\bibitem{hullthree}
C. M. Hull, JHEP {\bf 0111}, 012 (2001), arXiv:hep-th/0109213. 

\bibitem{ringstrom}
H. Ringstrom, Class. Quant. Grav. {\bf 17}, 713 (2000), arXiv:gr-qc/9911115; 
H. Ringstrom, Annales Henri Poincare {\bf 2}, 405 (2001), arXiv:gr-qc/0006035.
 
\bibitem{coley}
A. A. Coley, Class. Quant. Grav. {\bf 19}, L45 (2002), arXiv:hep-th/0110117. 

\bibitem{lidseycopeland}
E. J. Copeland, S-J. Lee, J. E. Lidsey, and S. Mizuno, 
Phys. Rev. D {\bf 71}, 023526 (2005), arXiv:astro-ph/0410110.
 
\bibitem{coleybook}
A. A. Coley, {\em Dynamical Systems and Cosmology} 
(Kluwer Academic Publishers, 2003).

\bibitem{bkl}
V. A. Belinskii, I. M. Khalatnikov, and 
E. M. Lifshitz, Sov. Phys. Usp. {\bf 13}, 745 (1971). 

\bibitem{dunsby}
P. K. S. Dunsby, N. Goheer, M. Bruni, and A. A. Coley, 
Phys. Rev. D {\bf 69}, 101303 (2004), arXiv:hep-th/0312174. 

\bibitem{coleylim}
A. A. Coley and W. C. Lim, 
Class. Quant. Grav. {\bf 22}, 3073 (2005), arXiv:gr-qc/0506097.
     


\end{thebibliography}
\end{document}